\documentclass[journal]{IEEEtran}
\usepackage[utf8]{inputenc}
\usepackage{color}
\usepackage{graphics}
\usepackage{graphicx}
\usepackage{amsmath}
\usepackage{ulem}

\usepackage{algorithm2e}

\newcommand{\fref}[1]{Fig.~\ref{#1}}

\newcommand{\tref}[1]{Table~\ref{#1}}

\title{Wireless 2.0: Towards an Intelligent Radio Environment Empowered by Reconfigurable Meta-Surfaces and Artificial Intelligence}
\author{Haris Gacanin and Marco Di Renzo}
\date{October 2019}

\begin{document}

\maketitle

\begin{abstract}
We introduce ``Wireless 2.0'': The future generation of wireless communication networks, where the radio environment becomes controllable, programmable, and intelligent by leveraging the emerging technologies of reconfigurable meta-surfaces and artificial intelligence (AI). This paper, in particular, puts the emphasis on AI-based computational methods and commence with an overview of the concept of intelligent radio environments based on reconfigurable meta-surfaces. Later we elaborate on data management aspects, the requirements of supervised learning by examples, and the paradigm of reinforcement learning (RL) to learn by acting. Finally, we highlight numerous open challenges and research directions.
\end{abstract}

\section{Introduction}
Next-generation wireless networks will be as pervasive as the air we breathe, not only connecting us but embracing us through a web of systems that support personal and societal well-being. The ubiquity, speed and low latency of such networks will allow currently disparate devices and services to become a distributed intelligent platform integrating communications, sensing, localization, and computing capabilities \cite{MDR_EURASIP}. This distributed platform, in particular, will possess perception, learning, reasoning, and decision-making capabilities, which will make artificial intelligence (AI) an indispensable tool to optimize and to efficiently operate it \cite{ai4coms}.

Recently, three fundamental technologies, namely small cells, massive multiple-input multiple-output (MIMO) systems and millimeter-wave (mmWave) communications, spearhead the emergence of the fifth-generation (5G) wireless networks. The question is, however, whether these technologies alone will be sufficient to build the distributed intelligent platform that next-generation wireless networks need \cite{MDR_Access}. %, \cite{MDR_Tx}. 
5G wireless networks, in addition, are rapidly evolving towards an intelligent and software-defined design paradigm, where different parts of the network might be configured and controlled via user-centric AI \cite{hg0}. In this optimization process, however, the radio environment -- the medium or propagation channel -- is generally assumed uncontrollable and often an impediment to be reckoned with.
 %For example, signal attenuation limits the network connectivity, multi-path propagation results in fading phenomena, reflections and refractions from objects are a source of uncontrollable interference.

Today, communication engineers are used to model the radio environment as an entity that the transmitters and receivers need to adapt to, in order to either counteract or to leverage the propagation channel. Typical approaches include multiple antennas, complex encoding/decoding, and advanced communication protocols. These approaches have allowed wireless networks to increase the capacity-per-unit-of-energy by a 1000-fold factor in the last 20 years \cite{MDR_EE}. However, contemporary wireless communication systems remain extremely inefficient due to the constraints imposed by the radio environment per se. A typical base station, for example, transmits radio waves of the order of magnitude of Watts while a typical user equipment detects signals of the order of magnitude of $\mu$Watts. The rest of the energy is either dissipated over the channel or is a source of interference for other network elements.

These fundamental limitations are challenged by recent research on intelligent radio environments (IREs) \cite{MDR_EURASIP}. In IREs, the technology enablers of reconfigurable meta-surfaces (RMSs) and AI-based data computational techniques are leveraged to enable the controllability, programmability, and optimization of the next-generation wireless networks. In this paper, the emerging vision of next-generation wireless is referred to as ``Wireless 2.0'' in light of the fundamental changes that it entails on current wireless communications and architectures. In the rest of this paper, we motivate, introduce, discuss, and overview the research challenges of Wireless 2.0 with focus on the application and implications of RSMs and AI-based computational methods.

\begin{figure*}[!t]
    \centering
    \includegraphics[width=6in]{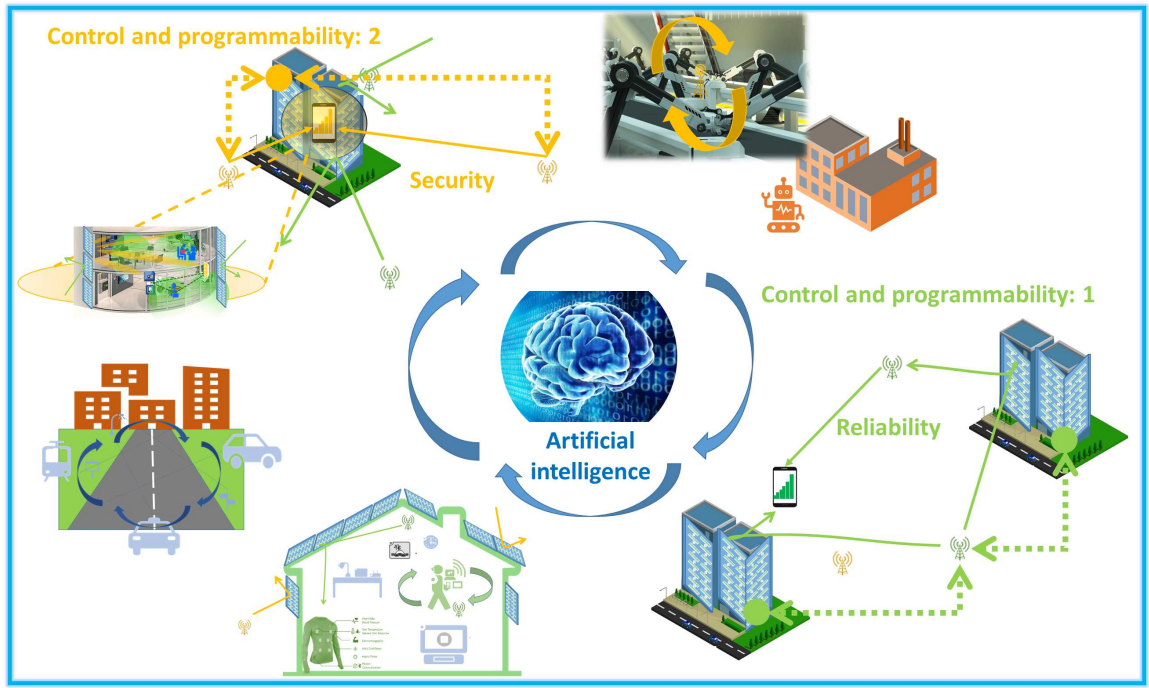}
    \caption{Examples of artificial intelligence application areas of intelligent radio environments with RMSs; higher reliability of communication connectivity, wireless security of high-profile enterprise clients.}
    \label{fig:apps}
\end{figure*}

\section{Wireless 2.0: From Adaptation to\\ Control and Programmability}
5G wireless networks are designed based on the fundamental postulate that only the end-points of the communication links, i.e., the transmitters and the receivers, can be optimized for improving the network performance. The propagation environment that lies in between them is, on the other hand, out of control of the communication engineers. In other words, while the transmitters and receivers can be programmed, controlled, and optimized, the environmental objects (e.g., walls, buildings, furniture, ceilings, floors, etc.) that constitute the wireless environment cannot be customized based on the network conditions.

This approach to design and optimize wireless networks has three fundamental limitations:
\begin{itemize}
  \item The ultimate performance limits of wireless networks may not have been reached yet. By jointly optimizing the transmitter, the receiver, and the environment, the performance of wireless networks may be further improved \cite{MDR_ISIT}. 
  \item In some application scenarios, the transmitters and receivers may not be made too complex. In the Internet of Things (IoT), for example, the devices are unlikely to be equipped with multiple antennas. Having the opportunity of customizing and controlling the environment may open new opportunities for network optimization \cite{MDR_MIT}.
  \item The radio waves are used inefficiently. When reflected or refracted by an object, for example, the energy is scattered towards unwanted directions, thus reducing the efficiency of utilization of the emitted power.
\end{itemize}

A paradigm-shifting wireless vision, which we refer to as Wireless 2.0 equips wireless networks with the functionalities of ($i$) customizing the radio environment (i.e. controlling the propagation of radio waves and programming the environmental objects to this end) besides the capability of optimizing the end-points of the communication links, and ($ii$) optimizing the resulting wireless communications with the aid of AI-based computational techniques.

\subsection{Wireless 2.0: Communications Empowered by Nearly-Passive Reconfiguable Meta-Surface}
The key technology for enabling communication engineers to customize the radio environment is constituted by the RMSs. An RMS is a thin sheet of electromagnetic material made of elementary elements (pixels), which are referred to as scattering particles or meta-atoms, that can be configured via external stimuli. Depending on the configuration of each individual pixel, the RMS is capable of altering the wavefront of a radio wave that impinges upon it. For example, an RMS can modify the direction of the reflected or refracted waves, the polarization of the scattered waves, or can modulate data onto the shape of the scattered waves. RMSs are, in particular, the two-dimensional equivalent of meta-materials, and are characterized by their very small thickness, which is much smaller than the wavelength of the radio waves. For this reason, RMSs are usually modeled as zero-thickness sheets of meta-materials. The two-dimensional nature of RMSs make them easier to design, less expensive, and easier to deploy than their three-dimensional counterpart. RMSs, therefore, are special surfaces that are engineered to possess properties that cannot be found in surfaces made of naturally occurring materials.

For their success and effective application to wireless networks, the RMSs need to have three main features:
\begin{itemize}
  \item \textit{Configurability}. In order to account for the dynamic nature of wireless environments, the RMSs need to dynamically adapt their response to the radio waves, after being manufactured and deployed. This can be realized either by distributing throughout the meta-surface low-power electronic circuits (diodes, varactors, etc.) that enable the RMSs to change their reponse according to the status of the electronic circuits, or by realizing the scattering particles that constitute the RMSs with reconfigurable material. The need of configurability increases the design complexity and cost of RMSs.
  
  \item \textit{Nearly-passive implementation}. In order to be cost-effective and not to further increase the carbon footprint of wireless networks, the RMSs need to be as passive as possible. This make them significantly different from conventional relays \cite{MDR_Relay}. Since the RMSs need to be configurable, it is unlikely, in practice, that they can be completely passive. The electronic circuits that enable their configurability consume, in fact, some power. The RMSs can be made, however, nearly-passive by using low power electronics and by employing energy harvesting modules, since no power amplifiers and signal processing capabilities are envisioned for their operation. Their nearly-passive implementation makes, however, challenging the design of protocols and algorithms to estimate the channel state information that is needed to program and control their operation \cite{MDR_CSI}. 
  
  \item \textit{Multi-function wave shaping}. In order to fully leverage the potential of RMSs, they need to be capable of realizing different functions at the same time, e.g., reflecting, refracting, and blocking the impinging radio waves. The design of configurable and multi-function RMSs is an open and timely research field \cite{MDR_Aalto}.
\end{itemize}

Broadly speaking, the overarching vision of employing RMSs in wireless networks consists of coating the physical objects that constitute the wireless environment with configurable meta-materials, which are capable of shaping the radio wave in arbitrary ways. Consequently, wireless networks are not designed anymore to adapt themselves to the environment, but the environment is part of the optimization space. In other words, the resulting wireless environment is not viewed as a random and uncontrollable entity anymore, but rather as part of the network design parameters that are subject to optimization in order to support diverse performance metrics, such as rate, latency, reliability, energy efficiency, privacy, and massive connectivity. Recent experimental results substantiating the feasibility of this vision are reported in \cite{MDR_PathLoss}.

\subsection{Wireless 2.0: Communications Empowered by AI}
\begin{figure}[!t]
    \centering
    \includegraphics[width=3in]{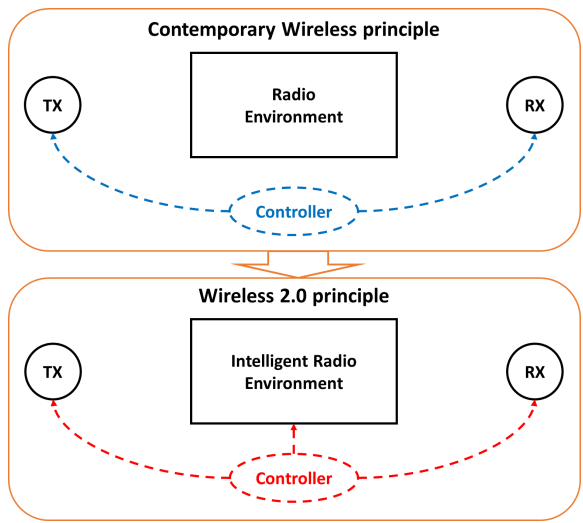}
    \caption{Transition from the contemporary wireless principle towards ``Wireless 2.0''.}
    \label{fig:wireless2.0}
\end{figure}

In order to optimize the operation of the RMSs, AI is viewed as an essential enabler. More precisely, the conceptual difference between contemporary Wireless and Wireless 2.0 is sketched in \fref{fig:wireless2.0}.  In Wireless 2.0, the communication system and the RMSs-enabled radio environment are jointly controlled by means of AI.  

\subsubsection*{Intelligent Radio Environments}
Wireless 2.0 is composed of three tightly coupled components: ($i$) the network elements (mobile terminals, base stations, etc.); ($ii$) the environmental objects coated with RMSs; and ($iii$) computational learning methods. These three components constitute the intelligent radio environments -- IREs.

RMSs-empowered wireless networks are a paradigm-shifting emerging concept. The design and optimization of such networks are, however, challenging. It is known that 5G wireless networks are already too complex for employing solely model-based methods to optimize their deployment, operation, and maintenance \cite{MDR_TCOM}. In other words, it is difficult to develop accurate and tractable models for the RMSs that account for their physics and electromagnetic nature and that, once plugged into wireless network are amenable for network optimization. AI provides an efficient approach for overcoming this issue and for leveraging the true potential of RMSs-empowered wireless networks. On the other hand, the computational complexity of deploying, programming and controlling RMSs-aided wireless networks rises significantly with with the increase of the network-to-infrastructure and user-to-network interactions. This requires more efficient and on-demand network intelligence to cope with complex deployment planning, real-time programmability for optimization, and dynamic control for service provisioning.

Recently, AI-based computational methods are an essential enabler for optimizing and operating 5G wireless networks. The possibility of customizing the propagation environment by deploying, programming, and controlling many RMSs that are distributed throughout the network makes AI-based computational methods essential to operate the resulting IREs. AI-enabled machines can be designed to perform ``intelligent'' tasks without being programmed to accomplish any single (repetitive) task, but adapting themselves to different environments. To this end, AI provides methods for designing network to autonomously interact with the environment in the way that humans consider intelligent – including all the characteristics of human cognitive abilities, i.e. planning, perceiving, reasoning, learning, and problem solving. AI defines a framework for knowledge manipulation (building new knowledge and exploiting already gained knowledge) through perception, reasoning (specifying what needs to be done, but not how) and acting \cite{ai4coms}. Today, applied AI (henceforth AI) based on machine learning (ML) methods are used to perform human cognitive abilities with focus on learning and decision-making driven by human-defined rules and constraints before ``actual learning''. The IREs, as a new component of the wireless network, need to be designed by leveraging and capitalizing on AI-based reasoning, acting, planning and learning \cite{ai4coms}.

\subsubsection*{Wireless 2.0 Controller}
Programmable wireless networks require full awareness of their complex and non-stationary environments. This can be described from the perspective of the IREs by how information is sent, transformed and observed as summarized in \tref{tab:transform}. %AI with ML defines an autonomous agent that performs a given task by improving its performance through learning and adapting to the environment from own experiences.
An AI-based controller for optimization of an IRE considers as an input the relative locations of the users, user mobility, relative locations of radio access nodes, the dynamics of network service demands, as well as the position, the distribution and the functions of RMSs. For example, glass antennas have been considered in 5G networks to cope with the propagation through windows in the millimeter-wave spectrum. Today, more than 50\% of the total surface of buildings in urban areas is made of glass. In the future, we may envision smart buildings whose glasses are made of RMSs that allow us to control, program, and optimize outdoor-to-indoor communications. Hence, the development of the emerging concept of IREs may be in symbiosis with the development of novel materials for the design of smart buildings in \textit{smart cities}, so as to support different requirements, e.g., reliable connectivity, security, as illustrated in \fref{fig:apps}. Other scenarios of application include \textit{smart homes}, \textit{autonomous driving}, \textit{factories of the future}, \textit{confined environments}, e.g., trains and airplanes, where the facades (e.g. made of glass) of buildings, walls, machines, and even clothing are coated with RMSs that can be programmed via external stimuli and optimized through AI. 

In the following sections, we elaborate on different aspects of AI as an essential enabler to realize the Wireless 2.0 vision, and motivate its need in the context of RMSs-empowered wireless networks. 

\begin{table}[t!]
  \begin{center}
    \caption{Exchange of partially observed information through intelligent radio environment.}
    \label{tab:transform}
    \begin{tabular}{|p{2.5cm}|p{2.5cm}|p{2.5cm}|}%{l|c|r} % <-- Alignments: 1st column left, 2nd middle and 3rd right, with vertical lines in between
      \hline
      \textbf{How message is sent} & \textbf{How message is received} & \textbf{How message is observed}\\\hline
      Based on the environment observed at the sender, actions are set. These actions are usually affected by noisy observations in the vicinity of the sender. Even with feedback, this provides limiting observation to the receiver. & The environment is partially observable and stochastic with an infinite horizon. This leads to highly noisy information at the output of the propagation environment. Furthermore, embedded sensors generate spatial-data with independent structures at different locations. & The signal is received from multiple sources at different locations. Thus, physical-layer sensors create different signatures for ``visualization'' at higher layers. Non-communication equipment such as cameras and microphones might be employed for sensing but integration of such information is challenged (ruled) by human logic. Through sensors receiver observes environment at its vicinity, which is a local interpretation of the environment with or without feedback from the sender. In addition to additive noise, the observation is distorted by 1) the propagation environment and 2) network-user interactions. \\
      \hline
    \end{tabular}
  \end{center}
\end{table}

\section{Learning as an Enabler of the Intelligent Radio Environments}
ML has been employed for several years to wireless networks to enhance decision-making by finding structures in data with ML methods – knowledge discovery – as a means to describe the system performance. Within the field of AI, ML evolved from computational learning theory as an efficient way to solve matching problems by processing and learning from given data with little (i.e. supervised) or no guidance at all (i.e. unsupervised) \cite{AIbook}. 

\begin{figure}
    \centering
    \includegraphics[width=.95\linewidth]{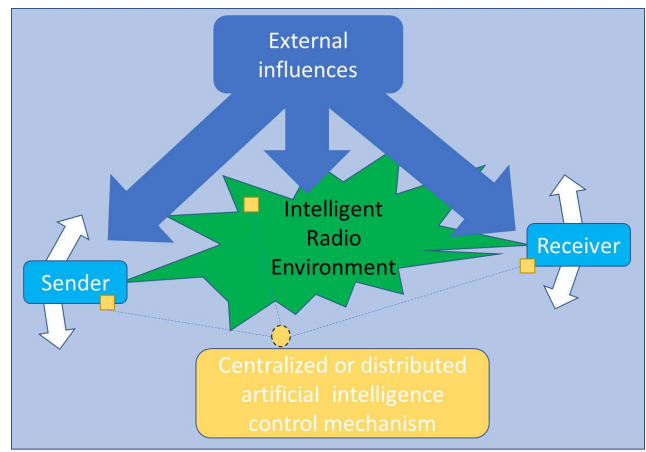}
    \caption{Principal elements of Wireless 2.0: 1) Sender/receiver pair, Centralized or distributed AI control mechanism, IRE and external influence from other networks.}
    \label{fig:ai-radio}
\end{figure}

An IRE with a mobile sender and receiver, a control system and the propagation environment is illustrated in \fref{fig:ai-radio}. Irrespective of the goal to predict or operate autonomously the AI-based control mechanism with learning might take different designs depending on the class of a learning method. Learning methods can be classified into training-based (i.e. supervised) being dependent on data management, and training-free (i.e. RL) being dependent on the availability of data in real time. The learning methods in IRE support a control process of modification of each RMS element to bring the sender, the environment and the receiver into an agreement given the feedback information in \fref{fig:ai-radio}. 

Next, we discuss data management, supervised and reinforcement learning requirements in IREs.

%In this paper, we do not attempt to give detailed description of the learning methods, but rather we refer readers to the references with examples \cite{hg1}.

\subsection{Data management} \label{sec:datamgmt}
Today wireless data is managed by deterministic collection rules, while several aspects need to be considered to support learning in IREs:

\subsubsection*{Adaptive control policy}
Since application clients can request different quality of service levels, data management needs to be carefully designed. This needs to be realized without generating large traffic in the network. Today, this is done via a collection of scheduled parameters that is constrained to network optimization in time given the location. For Wireless 2.0 we need an intelligent data controller to support real-time programmability for optimization. This may be accomplished by employing AI-based computational methods to gain and exploits knowledge about the dynamic (in space and time) service requirements and the evolution of the environment as illustrated in \fref{fig:ai-radio}. Finally, the data collection may be controlled in a non-intrusive fashion to avoid service interruptions leading to challenges with supervised learning. 

\subsubsection*{Amount of data and frequency of collection}
In the emerging IREs, the collection of different types of data needs to be controlled in correct amount and time intervals. The collection of too much or insufficient amount of data, as well as the collection of too frequent or too rare data may be detrimental for the operation of wireless network. This is in contrast with current approaches, where large-scale parameter collection is done for the purpose of prediction with learning models or devising insights from data by computational analytics \cite{hg0}. For example, an infrequent data collection limits root-cause analysis of IREs. This is because data has its own life-time. 

\subsubsection*{{Online data analytics}}
Recently, predictive data analytics has been employed to extract and transform the environmental data into actionable insights \cite{hg0}. The analytics identifies the state of the environment and closes the gap on insights into the wireless network and IRE. Thus, low collection response and processing time is a critical design parameter. For example, the proactive analytics may require to process structured or unstructured data by dynamic-scale, ultra-fast and low-power distributed data processing and storage technologies.

\subsubsection*{{Programmable data control}}
The data is shared by different elements in Wireless 2.0. Under the assumption that for the given problem/observation the data is spatially correlated, i.e. sensors at different locations capture the environmental data describing the same problem, we need to consider a programmable processing with spatial-temporal data. The data need to be flexible, scalable, and enriched with complementary information about different elements illustrated in \fref{fig:ai-radio}. How to recognize such data sources with current technology is a challenge. 

\subsection{Supervised learning -- Learning by examples}
Recently, deep learning has become an attractive supervised learning technique due to its performance superiority over traditional machine learning techniques (e.g. linear regression, decision trees) on problems including speech and vision recognition, natural language processing, and gaming. However, it is interesting to summarize design differences and advantages of both classical machine learning and deep learning in non-stationary environments.

%\textbf{Deep learning vs. traditional ML techniques}
\subsubsection*{Design of dataset and labelling}
For applications such as image and visual recognition, the testing accuracy with deep learning is significantly higher than classical ML techniques. Such high accuracy of deep learning is preceded by extremely large training (e.g. in millions of complex datasets). A large size of datasets may not always be available for training, while labelling and cleaning-up of the dataset is highly expensive, time-consuming and likely challenging for real time applications in non-stationary (e.g. mobile) environment. On the other hand, for applications with a smaller size of datasets, the accuracy of classical ML techniques is comparable or even higher than deep learning and their joint utilization is also possible.

\subsubsection*{Computational processing}
Today, deep learning relies on high-cost high-performance graphical processing units (GPUs) for efficient and reasonably fast training, assuming that labeled datasets are available. In most wireless applications at the physical and data link layers the GPUs on wireless nodes are not (yet) available. Thus, deep learning-enabled wireless devices with GPUs would require faster central processing units (CPUs) and larger solid-state drives, and very fast and large random access memory. Today, this is mostly reserved for scenarios where cloud computing is available. On the other hand, the training of classical ML techniques is faster and less expensive than deep learning allowing study on different techniques in a short period of time. In this case the training is possible with standard hardware, memory, and CPaUs.

\subsubsection*{Efficient data pre-processing}
The performance of classical ML techniques depends on the quality of feature engineering. With deep learning feature engineering is not required. Deep learning works directly on the input data conditioned that training is efficiently performed. The feature engineering for classical ML techniques may be a challenge for complex scenarios that lack domain expertise. On the other hand, feature engineering provides an effortless interpretation of the results. Another important point is the opportunity to adjust a set of parameters and the unambiguous redesign of the learning model. This is mainly because the relationship between the data and the learning model is explicit. In deep learning this is not the case since the model mimics “a black box” without need to understand how the data propagates throughout the deep learning model. In this case, adjustment of parameters set and redesign of learning model are a challenge due to the lack of theory. Finally, we mention that \textit{how}, \textit{when} and \textit{what} type of data to collect are highly critical design aspects.

%\h{Relate cross-points with challenges to enable smart environments!}

\subsubsection*{Transfer learning} 
In non-wireless applications, transfer learning allows effective utilization of a pre-trained deep learning model in different applications of the same domain. For instance, an image classification model may often be used as a feature extraction front-end to object detection and segmentation networks. This helps to reach high performance in a shorter period of time. To date, there has been only initial understanding of the applicability of transfer learning to different applications in wireless communications such as resource management, channel estimation, signal detection \cite{MDR_TCOM}. On the other hand, with classical ML techniques transfer learning is not possible because domain- and application-specific ML techniques with feature engineering are required to design learning models. The domain knowledge of different domains and applications is different and calls for specially designed study for each application.

\subsubsection*{When not to use deep learning}
There are some situations when deep learning driven control mechanism might not be suitable. If the response time is highly critical or depleted example data is available deep learning driven control mechanisms in IREs may create large errors and unacceptably long convergence time.

\begin{figure}
    \centering
    \includegraphics[width=.95\linewidth]{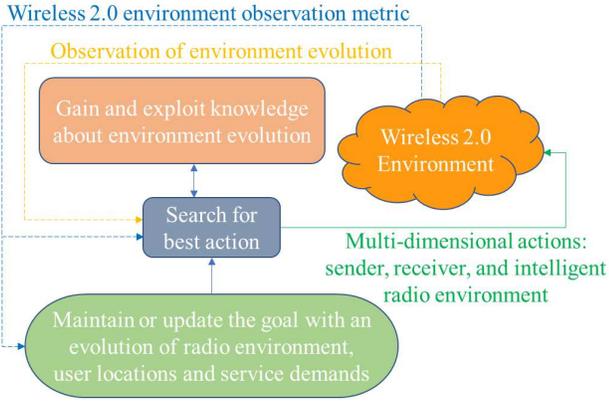}
    \caption{Learning to act: reinforcement learning principle.}
    \label{fig:rl}
\end{figure}

\subsection{Reinforcement learning -- Learning to act}
In the case of non-stationary environments we might consider another control mechanism for IREs: Reinforcement Learning.  RL is an online learning technique without need for data labelling and supervisor \cite{hg1}.

RL driven control mechanism receives a percept from the environment and decides to perform actions with a goal to search for a function that maps observations into actions in unknown environments as illustrated in \fref{fig:rl}. The control mechanism needs to perceive an IRE through a sequence of sensing, reasoning and acting to build its own experience in the form of knowledge base. The past experiences are employed to enhance new actions (e.g. good actions that achieve target quality-of-service are reused directly under similarly observed network conditions. On the other hand, bad actions with poor quality-of-service are used to refine the action search strategy. The two major RL designs are: 
\begin{itemize}
    \item A passive learner with a fixed policy that learns the values of actions and environment model;
    \item An active learner with a challenge to learn how to behave by environment/problem-specific design of exploration and exploitation mechanism.
\end{itemize}
The main differences between RL and supervised learning techniques are the following.

\subsubsection*{Online vs. offline decision-making}
The IREs in \fref{fig:ai-radio} need to support wireless communications in real time. Thus, control mechanism with online decision-making are required with fast optimization and adaptation. Unlike the offline decision-making of deep learning, a RL driven control mechanism introduces online decision-making to evaluate controller's new actions in non-stationary environments. This approach provides additional information and speed up the decision-making. On the other hand, RL driven control mechanism may exploit the currently available knowledge about the past actions and their state changes to make the best decision \cite{AIbook}. 

\subsubsection*{Integrated data collection and computational platforms}
Wireless 2.0 needs efficient data storage and data handling platforms. The design of data platforms needs to support both distributed and centralized controller mechanisms as illustrated in \fref{fig:ai-radio}. Data platforms, e.g. Hadoop, are already integrated with available learning tools such as TensorFlow, Scikit-learn, Microsoft Cognitive Toolkit. Such integration allows for the development of efficient supervised learning models with large-scale data manipulation running optimally on both central processing units and graphics processing units. However, in IREs in which the applications have strict latency and reliability constraints, fast-response data collection and processing are critical criteria. While being supported by a modest processing capabilities of handheld devices or battery-powered radio networking nodes, a development of flexible and fast-response data integration platforms need to be carefully designed.

\subsubsection*{Exploration -- Multi-dimensional action search}
RL-driven controller explores new actions to check if they can improve the system performance. Exploration employs closed-loop learning or heuristics/meta-heuristics due to problem complexity \cite{ai4coms}. Unlike contemporary exploration strategies, where new actions are selected according to a random selection policy (e.g. Boltzmann distribution), the IRE controller needs to drive action selection by environment-specific policy \cite{hg2}. Thus, semi-random exploration design without a negative impact on service delivery needs to be developed. Some examples of actions are the constituent elements of RMSs, sender/receiver adaptive modulation and coding, frame size and scheduling. Due to such complex action selection space, the exploration of probabilistic reasoning and inference based on belief or deep learning is an interesting research problem.

\subsubsection*{{Exploitation -- Experience/knowledge management}}
The controller needs to capture and exploit good experiences through interactions in IREs. By exploiting what it knows the controller maximizes immediate rewards greedily without trying new actions. The controller needs to build and exploit the knowledge to learn a (near-) optimal configuration action of RMSs by using a feedback mechanism (i.e. reward) as illustrated in \fref{fig:rl} \cite{hg2}. The strict control requirements of IREs described in previous section are supported by knowledge management of RL. For example, the exploration (optimizer) mechanism in IREs might have a negative effect on service delivery in real time \cite{hg1}. In fact, any changes of operating frequencies, sub-optimal deployment due to frequent mobility might result in a disruption of services \cite{hg2}. Due to the distributed nature of the problem, an efficient design of single or multiple instances of knowledge base is needed. Finally, we note that a domain knowledge base representation is application-specific.

% \subsubsection*{\textbf{Non-i.i.d. data sources}}
% While operating the intelligent radio environment the RL driven control mechanism gathers the relevant data described in \sref{sec:datamgmt}. However, the current action affects the data we observe in this moment, and consequently the future action selection becomes a function of current action. This leads to the critical design assumption that the data are not independent and identically distributed. This renders traditional statistical analysis methodology not applicable and requires different approaches to be devised.

\subsubsection*{When not to use reinforcement learning}
There are some situations when RL driven control mechanism might not be suitable, e.g. if the response time is not critical or abundant example data is available to drive supervised learning (either traditional or deep learning). RL driven control mechanisms in IREs, due to a large number of configurable elements (i.e. action space) of RMS might require a high computational processing and long convergence time.

\section{The Road Ahead}
This paper presented the concept of Wireless 2.0 emphasizing on AI computational methods as one of the enablers of IREs with RMSs. To this end, the solid understanding of traditional wireless communications models in combination with AI methods might lead us to principle designs of IREs. Nevertheless, this research opens numerous research directions and here we summarize the most prominent ones in the following three categories: data systems, learning mechanisms and knowledge management.

\subsubsection*{Data systems}
As discussed some of the AI computational methods rely on data introducing relevant concerns in the context of distributed data management and control, where required latency and reliability put constraint on feedback overhead. Thus, to devise fast processing and dynamic data architectures a study on predictive data analytics techniques for early radio environment sensing (i.e. diagnostics and troubleshooting) is needed either as distributed or federated mechanism. Consequently, a research on distributed data storage systems is required to support management of structured and unstructured data with dynamic-scale, ultra-fast and low-power distributed data control mechanisms.

\subsubsection*{Learning mechanisms}
Learning is an essential element of IREs. However, the performance limits of learning models in a non-stationary radio environment need to be understood first.
For example, it is not clear how to strike a balance between a speed of model training and evolution of non-stationary environment. Striking a balance highly depends on the application with design of architecture, computational method and available computational resources. To this end traditional machine learning techniques should not be discarded. It is important to understood under which conditions traditional techniques might be a better choice over deep learning techniques. In this case an important question is to automate feature engineering for traditional techniques, while dataset labeling and deep model training in non-stationary environments needs to be carefully studied. For example, design of deep RL with environment specific action selection mechanism is one interesting research study. On the other hand, if our environment-specific problem is defined by RL (i.e. Markov decision process), unlike traditional action selection by a random optimizer, we need to rethink the optimizer as a function of its environment while constrained to the dynamic service demand.

\subsubsection*{Knowledge management}
While machine learning techniques strictly devise learning mechanisms, an intelligent system is designed with broader disciplines of AI including decision-making, reasoning and knowledge management \cite{ai4coms}. For example, it is not clear what are the practical network-level performance trade-offs with AI and how to reach them. In another example, one might investigate how to dynamically design knowledge base to support highly accurate reasoning of the control mechanism. Optimization and automation of the knowledge base design for transfer learning across physically different environments is an interesting research direction. Finally, we are in need of clearly understanding of the potential performance gains and they economical justification, which can only be understood through adequate prototyping results.

\bibliographystyle{IEEEtran}

\begin{thebibliography}{99}

\bibitem{MDR_EURASIP} M. Di Renzo, \textit{et al.}, ``Smart radio environments empowered by reconfigurable AI meta-surfaces: An idea whose time has come'', EURASIP J. Wireless Commun. Networking, Vol. 129, 20 pages, May 2019.

\bibitem{ai4coms}
H. Gacanin, ``Autonomous Wireless Systems with Artificial Intelligence: A Knowledge Management Perspective,'' IEEE Vehicular Technology Magazine, Special issue on 6G: What is Next?, pp. 51 - 59, September 2019.

\bibitem{MDR_Access} 
E. Basar \textit{et al.}, ``Wireless communications through reconfigurable intelligent surfaces'', {IEEE Access}, Vol. 7, pp. 116753-116773, 2019.
	
%\bibitem{MDR_Tx} 
%W. Tang \textit{et al.}, `Wireless communications with programmable metasurface: New paradigms, opportunities, and challenges on transceiver design'',`\emph{ArXiv}, submitted. [Online]. Available: https://arxiv.org/pdf/1907.01956.pdf.
	
\bibitem{hg0}
H. Gacanin and M. Wagner, ``Artificial Intelligence Paradigm for Customer Experience Management in Next-Generation Networks: Challenges and Perspectives,'' IEEE Network Magazine, Vol. 33, Issue 2, March/April 2019. 




\bibitem{MDR_EE} A. Gati \textit{et al.}, ``Key technologies to accelerate the ICT green evolution: An operator’s point of view'', {IEEE Commun. Surveys Tuts.}, submitted. [Online]. Available: https://arxiv.org/pdf/1903.09627.pdf.


% \bibitem{MDR_SDN} C. Liaskos \textit{et al.},
% 	``A new wireless commun. paradigm through software-controlled metasurfaces'', {IEEE Commun. Mag.},
% 	vol. 56, no. 9, pp. 162-169, Sep. 2018.


\bibitem{MDR_ISIT} R. Karasik \textit{et al.},
	``Beyond max-SNR: Joint encoding for reconfigurable intelligent surfaces'', {ArXiv},
	submitted. [Online]. Available: https://arxiv.org/pdf/1911.09443.pdf.

	
\bibitem{MDR_MIT} V. Arun and H. Balakrishnan,
	``RFocus: Practical beamforming for small devices'', {ArXiv},
	submitted. [Online]. Available: https://arxiv.org/pdf/1905.05130.pdf.
	
	
% \bibitem{MDR_Orange} D.-T. Phan-Huy \textit{et al.},
% 	``Single-carrier spatial modulation for the Internet of Things: Design and performance evaluation by using real compact and reconfigurable antennas'', {IEEE Access}, vol. 7, pp. 18978-18993, 2019.

	
\bibitem{MDR_Relay} K. Ntotnin \textit{et al.},
	``Reconfigurable intelligent surfaces vs. relaying: Differences, similarities, and performance comparison'', {ArXiv},
	submitted. [Online]. Available: https://arxiv.org/pdf/1908.08747.pdf.
	
\bibitem{MDR_CSI} Z.-Q. He and X. Yuan,
	``Cascaded channel estimation for large intelligent metasurface assisted massive MIMO'', {ArXiv},
	submitted. [Online]. Available: https://arxiv.org/pdf/1905.07948.pdf.
	

\bibitem{MDR_Aalto} F. Liu \textit{et al.},
	``Intelligent metasurfaces with continuously tunable local surface impedance for multiple reconfigurable functions'', {Physical Review Applied}, vol. 11, no. 4, Apr. 2019.



\bibitem{MDR_PathLoss} W. Tang \textit{et al.},
	``Wireless communications with reconfigurable intelligent surface: Path loss modeling and experimental measurement'', {ArXiv},
	submitted. [Online]. Available: https://arxiv.org/pdf/1911.05326.pdf.
	

\bibitem{MDR_TCOM} A. Zappone \textit{et al.}, ``Wireless networks design in the era of deep learning: Model-based, AI-based, or both?'', {IEEE Trans. Commun.}, vol. 67, no. 10, pp. 7331-7376, Oct. 2019.

\bibitem{AIbook}
P. Norvig and S. J. Russell, \textit{Artificial Intelligence: A Modern Approach}, Prentice Hall, 3rd Edition, 2016.
	
\bibitem{hg1}
H. Gacanin \textit{et al.}, ``Self-Deployment of Non-stationary Wireless System by Knowledge Management with Artificial Intelligence,'' IEEE Trans. on Cognitive Communications and Networking, 2019.


\bibitem{hg2}
H. Gacanin \textit{et al.},``Self-optimization of Wireless Systems with Knowledge Management: An Artificial Intelligence Approach,'' IEEE Trans. on Vehicular Technology, Vol. 68, No. 10, pp. 9682-9697, October 2019.



% \bibitem{MDR_VTM} A. Zappone \textit{et al.},
% 	``Model-aided wireless artificial intelligence: Embedding expert knowledge in deep neural networks towards wireless systems optimization'', {IEEE Veh. Technol. Mag.},
% 	vol. 14, no. 3, pp.  60-69, Sep. 2019.



\end{thebibliography}

\end{document}